\documentclass[a4paper,final]{appolb}
\pdfoutput=1
\usepackage[pdftex]{graphicx}
\usepackage[unicode=true, bookmarks=false,bookmarksnumbered=true,bookmarksopen=true, breaklinks=false,backref=false,pagebackref=false, colorlinks=true]{hyperref}
\hypersetup{pdftitle={ },pdfauthor={ },linkcolor=black,citecolor=black, urlcolor=black, filecolor=black,pdfpagelayout=OneColumn,pdfnewwindow=true,pdfstartview=XYZ, plainpages=false}
\usepackage{color}
\usepackage{array}
\usepackage{tabularx}
\usepackage{booktabs}
\usepackage{enumerate}
\usepackage{verbatim}
\usepackage{bm}
\usepackage{amsmath}
\usepackage{amssymb}
\usepackage{algorithm,algorithmic}
\usepackage{pgf}
\usepackage{verbatim}
\usepackage{braket}
\usepackage[T1]{fontenc}
\usepackage[english]{babel}
\usepackage[english]{babel}
\usepackage{subfig}
\usepackage{float}
\usepackage{url}
\newcommand\T{\rule{0pt}{1.0ex}}
\newcommand\B{\rule[-0.8ex]{0pt}{0pt}}

\newcolumntype{Y}{>{\raggedright\arraybackslash}X}
\newcolumntype{W}{>{\raggedleft\arraybackslash}X}
\newcolumntype{Z}{>{\centering\arraybackslash}X}
\graphicspath{{figures/}}

\usepackage{verbatim}

\begin{document}
\title{Finite temperature lattice QCD with GPUs
\thanks{Presented at the International Meeting "Excited QCD", Les Houches, France, 20 - 25 February, 2011}}
\author{Nuno Cardoso, Marco Cardoso and Pedro Bicudo
\address{CFTP, Instituto Superior T\'ecnico\\ Avenida Rovisco Pais, 1, 1049-001 Lisboa, Portugal}}

\maketitle

\begin{abstract}
Graphics Processing Units (GPUs) are being used in many areas of physics, since the performance versus cost is very attractive.
The GPUs can be addressed by CUDA which is a NVIDIA's parallel computing architecture. It enables dramatic increases in computing performance by harnessing the power of the GPU.
We present a performance comparison between the GPU and CPU with single precision and double precision in generating lattice SU(2) configurations. 
Analyses with single and multiple GPUs, using CUDA and OPENMP, are also presented.
We also present SU(2) results for the renormalized Polyakov loop, colour averaged free energy and the string tension as a function of the temperature.
\end{abstract}

\section{Introduction}

Since the first release of CUDA (Compute Unified Device Architecture) by NVIDIA, the GPUs (Graphics Processing Units) are being addressed for physics computing in different areas where the performance is relevant. 
CUDA gives developers access to the GPU by virtual instruction set and memory of computational elements.  Whereas the CPU was projected for executing a single thread very quickly, the GPU architecture was projected to execute many concurrent threads slowly.

The most successful theories that describe elementary particle physics are the so called
gauge theories. SU(2) is an interesting gauge group, either to simulate the electroweak
theory, or to use as a simplified case of the SU(3) gauge group of the strong interaction.

However, generating SU(N) lattice configurations is a highly computationally demanding task and requires advanced computer architectures such as CPU clusters or GPUs. 

Nevertheless, GPUs are easier to access and maintain, as they can run on a local desktop computer, compared with CPU clusters.

This paper is divided in 3 sections. In section 2, we present the performance results and the results of the Polyakov loop, the colour averaged free energy and the string tension as well as a brief description how to calculate them. For a more detailed description on how to generate lattice SU(2) configurations in GPUs see \cite{Cardoso:2010di}. In section 3, we conclude.

\section{Results}

We implemented our code in CUDA language to run in one GPU or in several GPUs with OPEMMP. The code was tested in two different architectures, NVIDIA 295 GTX and NVIDIA 480 GTX cards, see Table \ref{tab:gpudef}.

\begin{table}[!htb]
\begin{center}
\small{
\begin{tabular}{|c|c|c|}
\hline
\T\B \small{\textbf{NVIDIA Geforce GTX}} & \textbf{295 (GT200)} & \textbf{480 (Fermi)}\tabularnewline
\hline
\T\B Number of GPUs & 2 & 1\tabularnewline
\hline
\T\B CUDA Capability & 1.3 & 2.0\tabularnewline
\hline
\T\B Number of cores & 2$\times$240 & 480\tabularnewline
\hline
\T\B Global memory & 896 MB per GPU & 1536 MB \tabularnewline
\hline
\T\B Number of threads per block & 512 & 1024\tabularnewline
\hline
\T\B Registers per block & 16384 & 32768\tabularnewline
\hline
\T\B Shared memory (per SM) & 16KB & 48KB or 16KB\tabularnewline
\hline
\T\B L1 cache (per SM) & None & 16KB or 48KB\tabularnewline
\hline
\T\B L2 cache (per SM) & None & 768KB\tabularnewline
\hline
\T\B Clock rate & 1.37 GHz & 1.40 GHz\tabularnewline
\hline
\end{tabular}}
\end{center}
\caption{NVIDIA's architecture specifications (SM means Streaming Multiprocessor).}
\label{tab:gpudef}
\end{table}

\subsection{Performance}
\begin{figure}[!htb]
\begin{centering}
    \subfloat[Single precision.\label{fig:sp}]{
\begin{centering}
    \includegraphics[width=0.28\paperwidth]{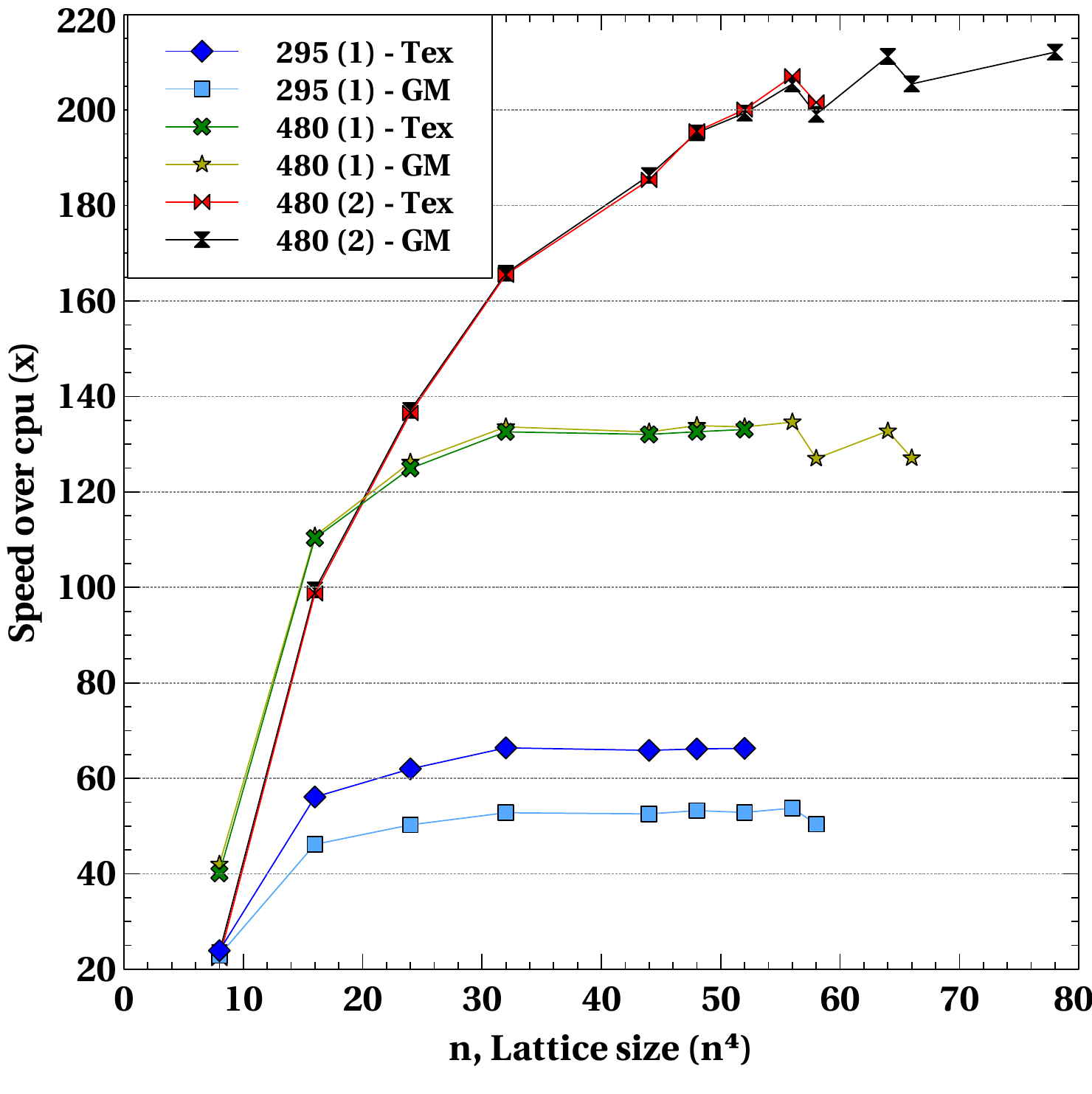}
\par\end{centering}}
    \subfloat[Double precision.\label{fig:dp}]{
\begin{centering}
    \includegraphics[width=0.28\paperwidth]{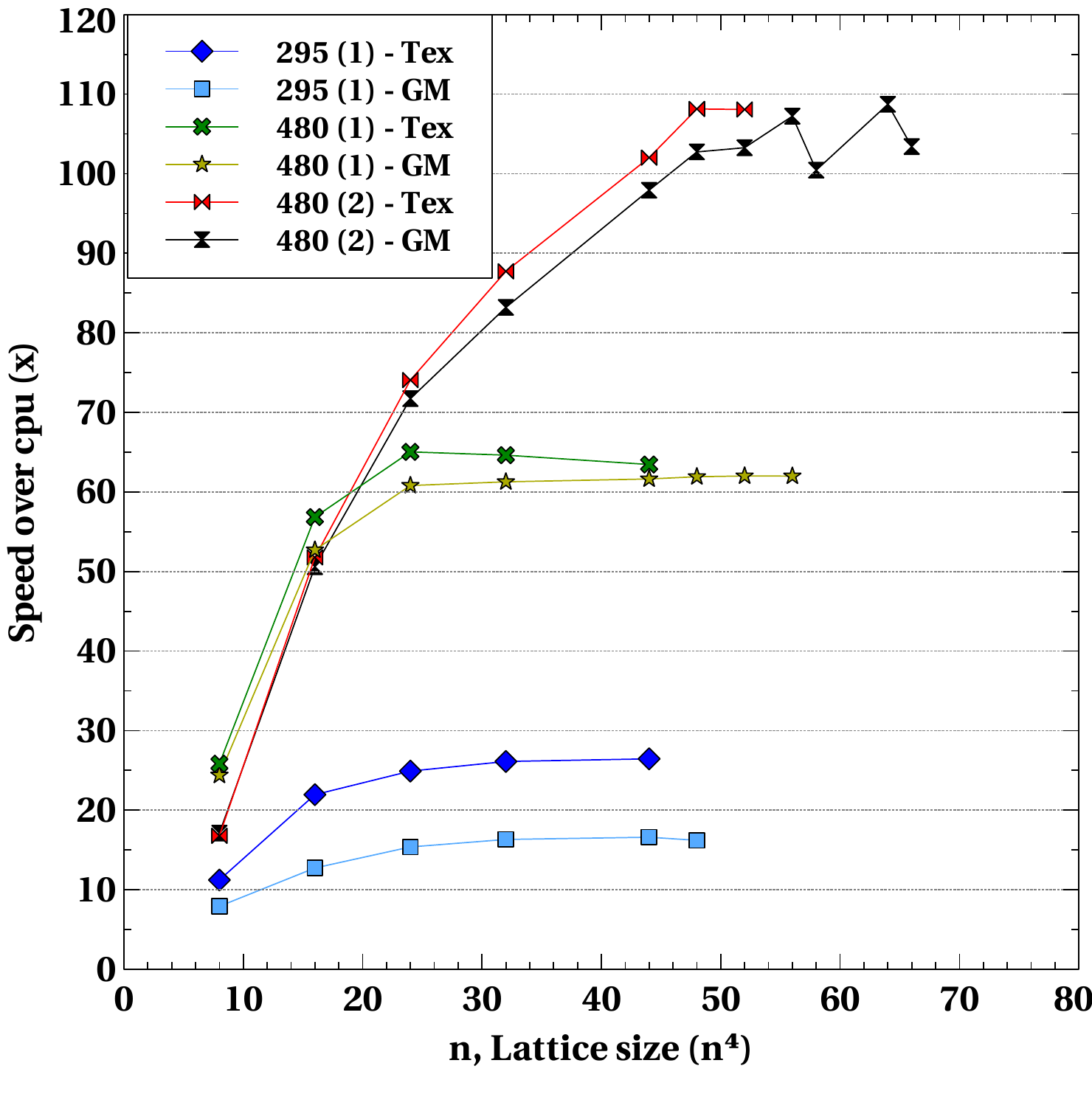}
\par\end{centering}}
\par\end{centering}
    \caption{Performance results. 295 - NVIDIA Geforce 295 GTX; 480 - NVIDIA Geforce
480 GTX; (1) - with 1 GPU; (2) - with 2 GPUs; Tex - using textures; GM - using global
memory.}
    \label{fig:performance}
\end{figure}

In order to test the GPU performance, we measure the execution time for the CUDA code
implementation in one, two GPUs and the serial code in one CPU core (CPU Intel$^{\text{(R)}}$ Core$^{\text{(TM)}}$ i7 CPU 920, 2.67GHz, 8 MB of L2 Cache and 12GB of RAM) for different lattice sizes at $\beta = 6.0$ with random SU(2) matrix initialization followed by 100 iterations of the heat bath method and the calculation of the mean average plaquette at each iteration, see Fig. \ref{fig:performance}.
For a more detailed overview see \cite{Cardoso:2010di}.

\subsection{Finite Temperature}

The Polyakov loop, $\Braket{L}$, is an order parameter for the deconfinement transition, \cite{McLerran:1981pb}, it measures the free energy, $F_q$, of a single static quark at temperature $T$,
\begin{equation}
\Braket{L} \propto \exp\left(-\frac{F_q}{T}\right)
\end{equation}
 where $T$ is connected to the lattice spacing $a$ by $T=1/(aN_t)$.
 
\begin{figure}
\begin{centering}
    \subfloat[\label{fig:FT_PL}]{
\begin{centering}
    \includegraphics[width=0.37\paperwidth]{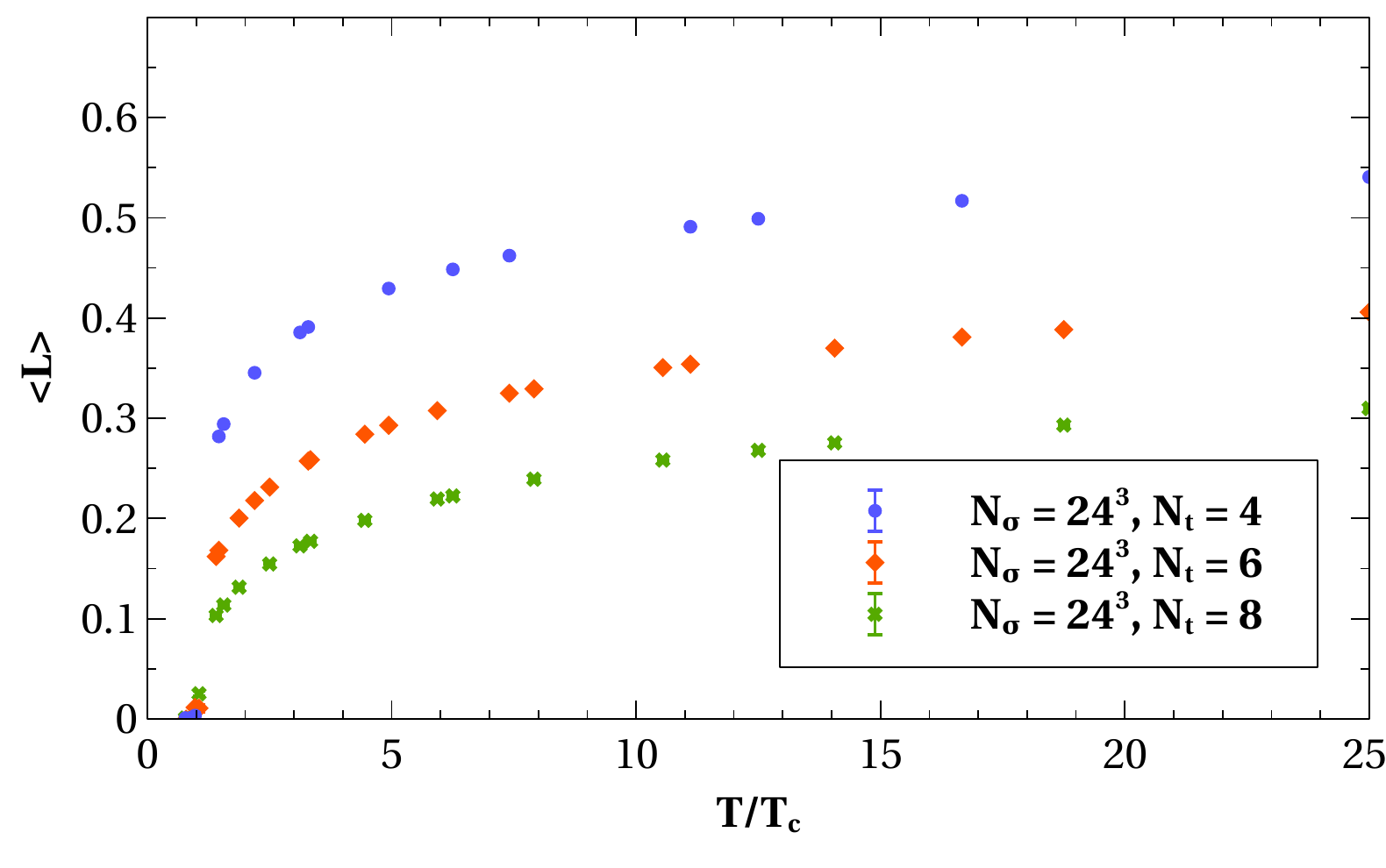}
\par\end{centering}}

    \subfloat[\label{fig:FT_PLr}]{
\begin{centering}
    \includegraphics[width=0.35\paperwidth]{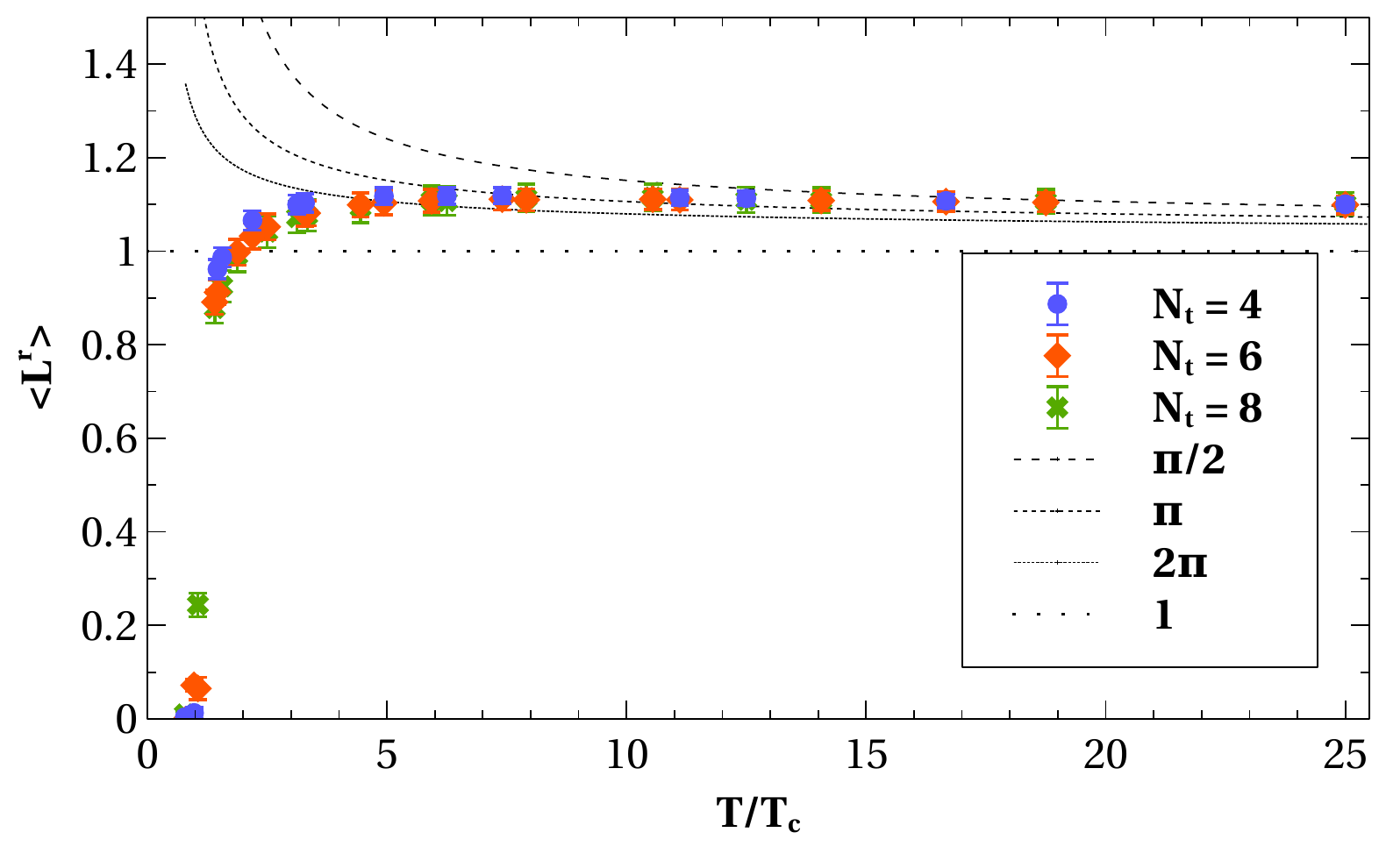}
\par\end{centering}}
\par\end{centering}
    \caption{Polyakov loop. \subref{fig:FT_PL} $-$ unrenormalized SU(2) Polyakov loop, $\Braket{L}$, at finite temperature. \subref{fig:FT_PLr} $-$ SU(2) Polyakov loop renormalized, the dotted lines correspond to the pure gauge Polyakov loop in HTL perturbation theory for $\pi/2,\ \pi,\ 2 \pi$.}
    \label{fig:FT_PL_all}
\end{figure}

The results for the Polyakov loop, Fig. \ref{fig:FT_PL}, show a dependence on the extension of the lattice in time direction. This is due to the self-energy contribution of the static quark source used as order parameter. 

Elimination of this self energy term is necessary to obtain an order parameter which is a function of the temperature alone. 

This can be done using the renormalization procedure described in \cite{Gupta:2007ax} and using the values of \cite{Bali:1993tz} obtained for the effective potential as the seed values. The renormalized Polyakov loop can be written as
\begin{equation}
\Braket{L^r} = \left(Z(g^2)\right)^{N_t} \Braket{L}
\end{equation}
where the renormalization constants $Z(g^2)$ should only depend on the bare coupling and fitting the values of $Z(g^2)$ obtained with this procedure with $Z(g^2)=\exp\left(A\,g^2+B\,g^4\right)$, we obtain $A=0.0637(18)$ and $B = 0.0731(16)$ with $\chi^2/dof= 1.16613$ for $g^2<1.3$. Applying this last results to all of our results in Fig. \ref{fig:FT_PL}, we obtain a renormalized Polyakov loop, Fig. \ref{fig:FT_PLr}, which is independent of the extension of the lattice in the time direction.
At high temperatures, the renormalized Polyakov loop approachs their corresponding HTL result.

\begin{figure}
\begin{centering}
    \subfloat[\label{fig:F_avg}]{
\begin{centering}
    \includegraphics[width=0.41\paperwidth]{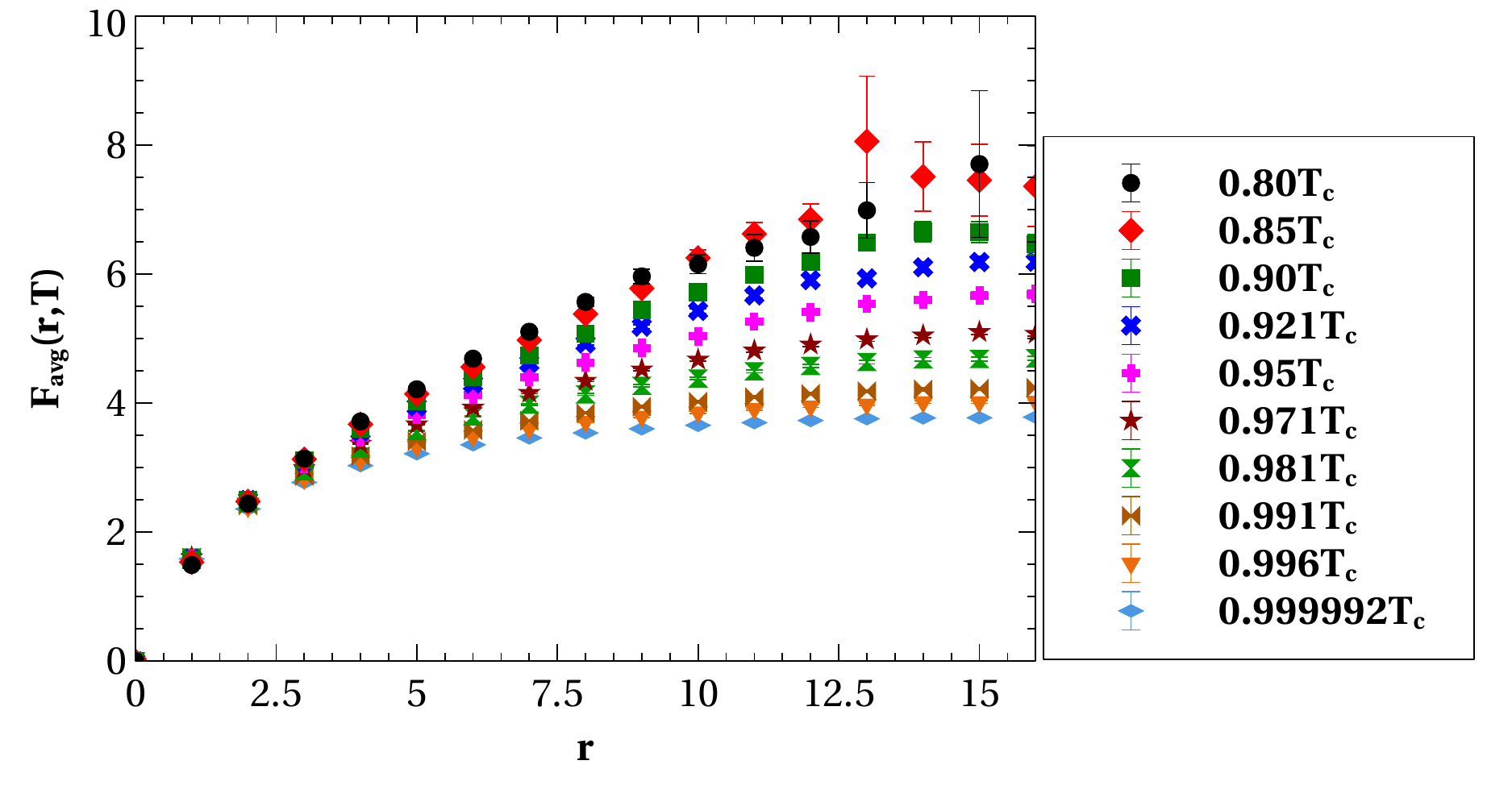}
\par\end{centering}}

    \subfloat[\label{fig:sigmaT}]{
\begin{centering}
    \includegraphics[width=0.37\paperwidth]{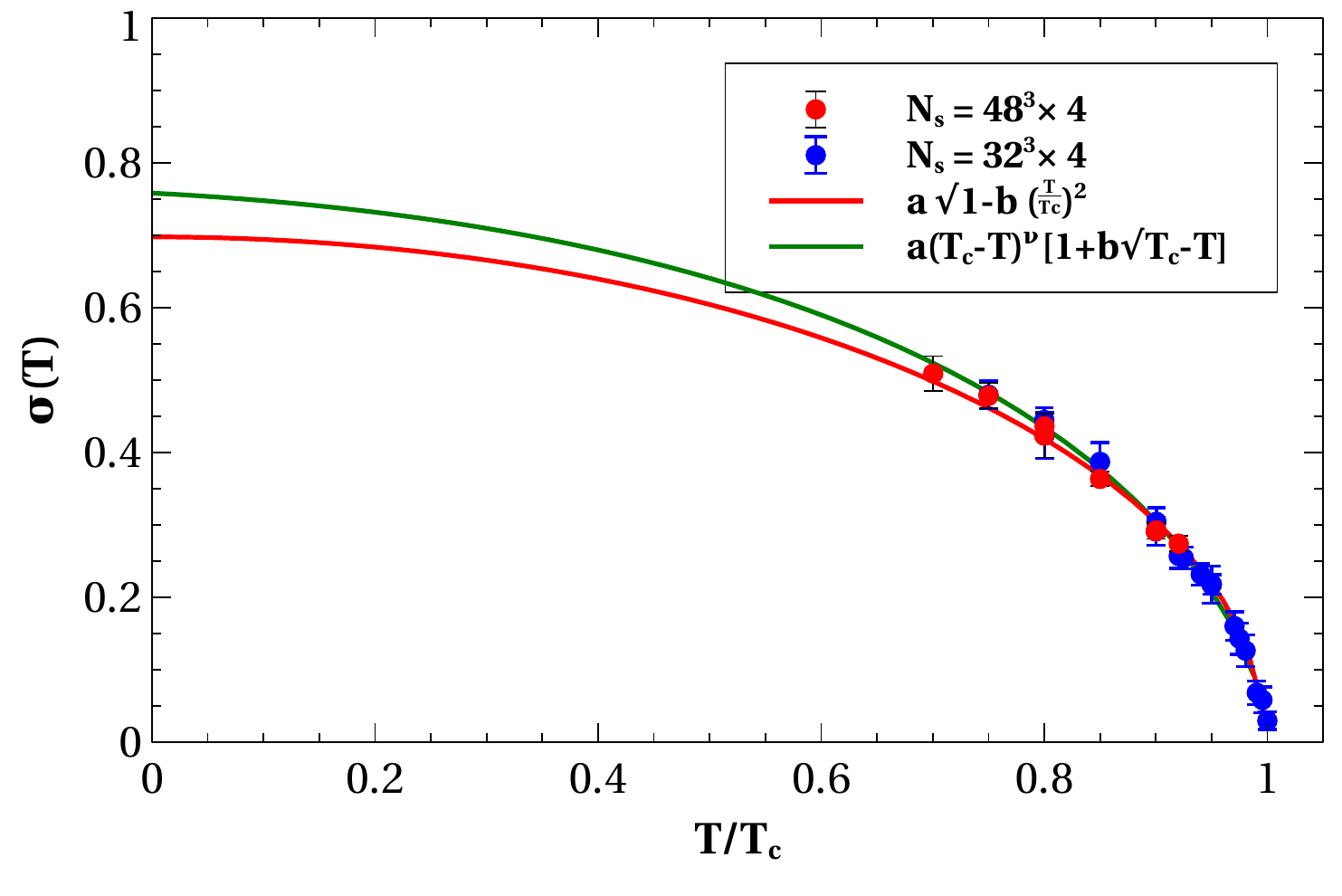}
\par\end{centering}}
\par\end{centering}
    \caption{\subref{fig:F_avg} $-$ SU(2) color averaged free energy, $F_\text{avg}$. \subref{fig:sigmaT} $ - $ SU(2) string tension, $\sigma(T)$.}
    \label{fig:FT_PL_all}
\end{figure}

The colour averaged free energy is defined as the correlation between two Polyakov loops,
\begin{equation}
\e^{-F_{\text{avg}}(r,T)/T+C}=\frac{1}{4}\Braket{\Tr L(y)\Tr L^\dagger(x)}
\end{equation}
which is gauge invariant. To eliminate the trivial temperature dependence due to the colour trace normalization, we apply $F_{\text{avg}}(r,T) \rightarrow F_{\text{avg}}(r,T) - T \ln 4$.
Fitting the $F_\text{avg}(r,T)$ data in Fig. \ref{fig:F_avg} with $F_\text{avg}(r,T)$ with $a_0 (T) - \frac{a_1(T)}{r} + \sigma(T) r$, we show in Fig. \ref{fig:sigmaT} the results for $\sigma(T)$ as a function of the temperature. 
Although the string tension in SU(2) was already addressed by \cite{Digal:2003jc}, the number of data points is too low to have a clear overview.
We fit our results with two different ansatz, $a \sqrt{1-b\left(T/T_c\right)^2}$ and $a\, (T_c - T)^\nu [1 + b\, \sqrt{T_c - T}]$ and obtain a reasonable $\chi^2/dof$ for the both fits. For the first ansatz, we obtain $a = 0.6976 \pm 0.0176$, $b = 0.9990 \pm 0.0059$ and $\chi^2/dof = 0.732$. In the second, we fix $\nu = 0.63$ according the 3D Ising exponent for the correlation length and obtain $a = 1.5541 \pm 0.0435$, $b = -0.5122 \pm 0.0576$ and $\chi^2/dof = 0.598$. Nevertheless, we need more data for $T < 0.7 T_c$.

\section{Conclusions}
With 2 NVIDIA GTX 480, we were able to obtain more than 200$\times$ the performance over one CPU core in single precision.
It's not possible to generate SU(2) configurations using only the GPU shared memory due to the limited amount of shared memory available. 
The limited number of registers also affects the GPU performance. 
Using texture memory in this problem, we were able to achieve high performance, both in the GPU without cache memory and in the GPUs with cache memory. 
However, in the GPUs with cache memory the difference is bigger in double precision than in single precision.
The occupancy and performance of the GPUs is strongly connected to the number of threads per block, registers per thread, shared memory per block, memory access, read and writing, patterns.
To maximize performance it is necessary to ensure that the memory access is coalesced and to minimize copies between GPU and CPU memories.

The renormalized Polyakov loop for $N_t\geq 4$ shows very small dependence on the lattice time direction for $N_t=4$ and low $T$. The string tension as a function of the temperature, $\sigma(T)$, extracted from the colour averaged free energy, for two different spatial lattice sizes does not reveal any volume dependence. The string tension for $T> T_c$ is zero, however for $T < T_c$ is temperature dependent.
We fit the string tension with two different ansatz, however, we need more data for $T< 0.7 T_c$. 
Future work will be dedicated to the study of this case.

\section*{Acknowledgments}
This work was financed by the FCT contracts POCI/FP/81933/2007, CERN/FP/83582/2008, PTDC/FIS/100968/2008 and CERN/FP/109327/2009.
Nuno Cardoso is also supported by FCT under the contract SFRH/BD/44416/2008.


\end{document}